\documentclass{article}
\usepackage{graphicx}
\usepackage{amsmath}

\input{tcilatex}

\begin{document}

\author{S. Manoff \\
\textit{Bulgarian Academy of Sciences}\\
\textit{\ Institute for Nuclear Research}\\
\textit{\ and Nuclear Energy}\\
\textit{\ Department of Theoretical Physics}\\
\textit{\ Blvd. Tzarigradsko chaussee 72}\\
\textit{\ 1784 Sofia - Bulgaria}}
\title{Deviation equations of Synge and Schild over $(\overline{L}_n,g)$-spaces }
\date{\textit{e-mail address: smanov@inrne.bas.bg}}
\maketitle

\begin{abstract}
Deviation equation of Synge and Schild has been investigated over
differentiable manifolds with contravariant and covariant affine connections
(whose components differ not only by sign) and metric [$(\overline{L}_n,g)$%
-spaces]. It is shown that the condition $\pounds _\xi u=0$ for obtaining
this equation is only a sufficient (but not necessary) condition. By means
of a non-isotropic (non-null) vector field $u$ [$g(u,u)=e\neq 0$] and the
metric $h_u$, orthogonal to it, a projected deviation equation of Synge and
Schild has been obtained for the orthogonal to $u$ vector field $\xi _{\perp
}$ and its square $L^2=g(\xi _{\perp },\xi _{\perp })$. For a given
non-isotropic, auto-parallel and normalized vector field $u$ this equation
could have some simple solutions.

\end{abstract}

\section{Introduction}

In the last decades the models of space-time have been generalized from
(pseudo) Riemannian spaces without torsion (denoted $V_n$-spaces) or with
torsion (denoted $U_n$-spaces) to spaces with contravariant and covariant
affine connections (whose components differ only by sign) and metrics
[denoted $(L_n,g)$-spaces] as well as to spaces with contravariant and
covariant affine connections (whose components differ not only by sign)
[denoted $(\overline{L}_n,g)$-spaces] \cite{Manoff-0}. It has been proved
that in these spaces the principle of equivalence holds \cite{Iliev-1}-\cite
{Iliev-4}, \cite{Hartley} and special type of transports (called
Fermi-Walker transports) \cite{Manoff-01}, \cite{Manoff-01a} exist which do
not deform a Lorentz basis. Therefore, the law of causality is not abuse in $%
(L_n,g)$- and $(\overline{L}_n,g)$-spaces if one uses a Fermi-Walker
transport instead of a parallel transport (used in a $V_n$-space). Moreover,
there also exist other types of transports (called conformal transports) 
\cite{Manoff-02}, \cite{Manoff-02a} under which a light cone does not
deform. At the same time, the auto-parallel equation can play the same role
in $(L_n,g)$- and $(\overline{L}_n,g)$-spaces as the geodesic equation does
in the Einstein theory of gravitation (ETG) \cite{Manoff-g1}, \cite
{Manoff-g2}. On this basis, many of the differential-geometric constructions
used in the ETG in $V_4$-spaces could be generalized for the cases of $%
(L_n,g)$- or $(\overline{L}_n,g)$-spaces. Bearing in mind this background a
question arises about applications of generalizations of well constructed
mathematical models in the ETG to theories in $(L_n,g)$- and $(\overline{L}%
_n,g)$-spaces. Such models, for instance, are deviation equations used as
theoretical basis for construction of gravitational wave detectors in ETG.
They can be generalized for $(L_n,g)$- and $(\overline{L}_n,g)$-spaces and
are worth being investigated.

The task of this paper is to show that deviation equations can be used in
the same way as in the ETG in gravitational theories using $(\overline{L}%
_n,g)$-spaces as a model of space-time. \textit{Deviation equations are
independent of the gravitational equations} conditions for finding out the
relative acceleration (as kinematic characteristic) between moving particles
in space with affine connections and metrics. Gravitational equations of the
type of those in the ETG (as dynamic characteristics) impose only additional
conditions on the curvature tensor and at the same time they give the
explicit form of these quantities for the corresponding theoretical model.

In the present paper the deviation equation of Synge and Schild is
generalized for $(\overline{L}_n,g)$-spaces and specialized for description
of the variation of the second covariant derivative of a vector field $\xi $%
, orthogonal to the non-isotropic (non-null) (time like for $n=4$) vector
field $u$. The vector field $\xi $ is interpreted as a deviation vector. In
Sec. 3. the generalized deviation equation of Synge and Schild and its
projective form (projective deviation equation of Synge and Schild) are
considered in $(\overline{L}_n,g)$-spaces. An analogous deviation equation
for the square of a non-isotropic vector (which is space like for $n=4$) is
found and investigated. In Sec. 4. the projective deviation equation of
Synge and Schild for the square of an auto-parallel $(\nabla _uu=0)$
(non-isotropic) and normalized [$g(u,u)=e=const.\neq 0$] vector field $u$ in 
$(\overline{L}_n,g)$-spaces as well as in $\overline{U}_n$-, and $\overline{V%
}_n$-spaces [as special cases of $(\overline{L}_n,g)$spaces] is considered.
Some simple solutions are found and examples for the case of $\overline{V}_n$%
-spaces are given which can lead to an equation in the form of an oscillator
equation. The results can be easily specialized for canonical contraction
operator $C$ in (pseudo) Riemannian spaces with and without torsion ($U_n$-
and $V_n$-spaces).

\section{Deviation equations}

In the general relativity, as a basis for the theoretical scheme for
gravitational wave detectors proposed by (Weber 1958-1961) and discussed by
many authors (Zacharov 1972), (Amaldi, Pizzella 1979), (Will 1979, 1981),
(Bicak, Rudenko 1987), the geodesic deviation equation (proposed by
Levi-Civita in 1925 in a co--ordinate basis) \cite{Ciufolini} in the form 
\begin{equation}
\frac{D^{2}\xi ^{i}}{ds^{2}}=R^{i}\text{ }_{jkl}u^{j}u^{k}\xi ^{l}\text{ , \
\ \ \ \ \ \ \ }u^{i}\text{ }_{;j}u^{j}=a^{i}=0\text{ ,}  \label{1}
\end{equation}

\noindent or in the index free form 
\begin{equation}
\nabla _{u}\nabla _{u}\xi =[R(u,\xi )]u\text{ , \ \ \ \ \ }a=\nabla _{u}u=0%
\text{ ,}  \label{2}
\end{equation}

\noindent has been used. Its generalization for non-geodesic trajectories ($%
a\neq 0$) (proposed by Synge and Schild in 1956 in a co-ordinate basis) in
the form 
\begin{equation}
\frac{D^{2}\xi ^{i}}{ds^{2}}=R^{i}\text{ }_{jkl}u^{j}u^{k}\xi ^{l}+a^{i}%
\text{ }_{j}\xi ^{j}\text{ , \ \ \ \ \ \ \ \ }a^{i}=u^{i}\text{ }_{;j}u^{j}%
\text{ ,}  \label{3}
\end{equation}

\noindent or in index free form 
\begin{equation}
\nabla _u\nabla _u\xi =[R(u,\xi )]u+\nabla _\xi a\text{ ,}  \label{4}
\end{equation}

\noindent has also been used by Weber in a special form for construction of
gravitational waves detectors of the type of massive cylinders reacting to
periodical gravitational processes. The application of these equations in
experiments for detecting gravitational waves turned the attention of many
authors to considerations and proposals for new deviation equations.

From mathematical point of view many of the proposed by different authors
deviation equations can be obtained from the s.c. generalized deviation
identity (generalized deviation equation) in $(L_n,g)$-spaces \cite{Manoff-d}
- \cite{Manoff-d3} 
\begin{equation}
\nabla _u\nabla _u\xi \equiv [R(u,\xi )]u+\nabla _a\xi +T(\xi ,a)-\nabla
_u[T(\xi ,u)]+[\pounds _\xi \Gamma (\xi ,u)]u\text{ ,}  \label{5}
\end{equation}

\noindent or in a (co-ordinate or non-co-ordinate) basis 
\begin{equation}
\begin{array}{c}
(\xi ^i\text{ }_{;j}u^j)_{;k}u^k\equiv R^i\text{ }_{klj}u^ku^l\xi ^j+\xi ^i%
\text{ }_{;j}a^j+T_{kl}^i\xi ^ka^l-(T_{kl}^i\xi ^ku^l)_{;j}u^j+ \\ 
\\ 
+\pounds _\xi \Gamma _{kl}^i.u^k.u^l\text{ ,}
\end{array}
\label{6}
\end{equation}

\noindent where the components $\pounds _{\xi }\Gamma _{jk}^{i}$ are the Lie
derivatives of the components $\Gamma _{jk}^{i}$ of the contravariant affine
connection $\Gamma $ and 
\begin{equation}
\begin{array}{c}
a=\nabla _{u}u=u^{i}\text{ }_{;j}u^{j}.e_{i}=a^{i}.e_{i}\text{ , \ \ \ \ \ }%
u\in T(M)\text{ ,} \\ 
e_{i}=\partial _{i}=\partial /\partial x^{i}\text{ (in a co-ordinate basis) ,%
} \\ 
u^{i}\text{ }_{;j}=e_{j}u^{i}+\Gamma _{kj}^{i}u^{k}\text{ , \ \ \ \ \ \ \ \
\ \ \ \ }\Gamma _{kj}^{i}\neq \Gamma _{jk}^{i}\text{ ,}
\end{array}
\label{7}
\end{equation}

The operator $R(\xi ,u)$ is the curvature operator 
\begin{equation}
\begin{array}{c}
R(\xi ,u)=-R(u,\xi )=\nabla _\xi \nabla _u-\nabla _u\nabla _\xi -\nabla
_{\pounds _\xi u}= \\ 
\\ 
=[\nabla _\xi ,\nabla _u]-\nabla _{[\xi ,u]}\text{ , }\xi ,u\in T(M)\text{ ,}
\end{array}
\label{8}
\end{equation}

The operator $\pounds \Gamma (\xi ,u)$ is the deviation operator \cite
{Manoff-d1} 
\begin{equation}
\begin{array}{c}
\pounds \Gamma (\xi ,u)=\pounds _\xi \nabla _u-\nabla _u\pounds _\xi -\nabla
_{\pounds _\xi u}= \\ 
\\ 
=[\pounds _\xi ,\nabla _u]-\nabla _{[\xi ,u]}\text{ , \thinspace \thinspace
\thinspace \thinspace \thinspace \thinspace \thinspace \thinspace \thinspace
\thinspace \thinspace \thinspace \thinspace \thinspace \thinspace \thinspace
\thinspace \thinspace \thinspace \thinspace }\xi ,u\in T(M)\text{ ,}
\end{array}
\label{9}
\end{equation}

$\pounds _\xi u$ is the Lie derivative of the contravariant vector field $u$
along the contravariant vector field $\xi ,$%
\begin{equation}  \label{10}
\pounds _\xi u=[\xi ,u]=\nabla _\xi u-\nabla _u\xi -T(\xi ,u)\text{ ,}
\end{equation}

$\nabla _u\xi $ is the covariant derivative of the vector field $\xi $ along
the vector field $u,$ $T(\xi ,u)$ is the torsion vector field 
\begin{equation}
\begin{array}{c}
T(\xi ,u)=T_{kl}^i.\xi ^k.u^l.e_i\text{ ,} \\ 
\\ 
T_{kl}^i=\Gamma _{lk}^i-\Gamma _{kl}^i\text{ (in a co-ordinate basis }%
\{\partial _i\}\text{),} \\ 
\\ 
T_{kl}^i=\Gamma _{lk}^i-\Gamma _{kl}^i-C_{kl}\text{ }^i\text{ (in a
non-co-ordinate basis }\{e_i\}\text{),} \\ 
\\ 
\pounds _{e_k}e_l=[e_k,e_l]=C_{kl}\text{ }^i.e_i\text{ .}
\end{array}
\label{11}
\end{equation}

Since the deviation equations are related to the second covariant derivative
of a deviation vector $\xi $ they could be represented by means of the
kinematic characteristics related to the notions of relative accelerations
(shear, rotation and expansion accelerations) \cite{Manoff-05} and to their
corresponding relative velocities (shear, rotation and expansion
velocities). In Einstein's theory of gravitation notions such as shear
(shear velocity) $\sigma ,$ rotation (rotation velocity) $\omega $ and
expansion (expansion velocity) $\theta $ are used for invariant
classification of solutions of the Einstein's field equations. These notions 
\cite{Kramer} can be defined for vector fields over $\overline{(L}_n,g)$%
-spaces in analogous way as in $V_n$- and $U_n$-spaces by means of
representation of the covariant derivative of a vector field $\xi $ along
(another) non-isotropic (non-null) vector field $u$ [$g(u,u)=e\neq 0$] in
the form \cite{Manoff-05} 
\begin{equation}
\begin{array}{c}
\nabla _u\xi \equiv \frac 1e.g(u,\nabla _u\xi ).u+\overline{g}[h_u(\frac
le.a-\pounds _\xi u)]+ \\ 
\\ 
+\overline{g}[\sigma (\xi )]+\overline{g}[\omega (\xi )]+\frac 1{n-1}.\theta
.\overline{g}[h_u(\xi )]\text{ ,}
\end{array}
\label{12}
\end{equation}

\noindent or in a given basis in the form 
\begin{equation}
\begin{array}{c}
\xi ^i\text{ }_{;j}u^j\equiv \frac 1e.g_{\overline{k}\overline{l}}.u^k.\xi ^l%
\text{ }_{;m}u^m.u^i+g^{ij}[h_{\overline{j}\overline{k}}(\frac
le.a^k-\pounds _\xi u^i)+ \\ 
\\ 
+(\sigma _{\overline{j}\overline{k}}+\omega _{\overline{j}\overline{k}%
}+\frac 1{n-1}.\theta .h_{\overline{j}\overline{k}})\xi ^k]\text{ ,}
\end{array}
\label{13}
\end{equation}

\noindent where 
\begin{equation}
\begin{array}{c}
h_u=g-\frac 1e.g(u)\otimes g(u)=h_{ij}.e^i.e^j\text{ , }g=g_{kl}.e^k.e^l%
\text{ ,} \\ 
\\ 
e^i.e^j=\frac 12(e^i\otimes e^j+e^j\otimes e^i)\text{ ,} \\ 
e^i=dx^i\text{ (in a co-ordinate basis) ,} \\ 
\\ 
h_u(u)=h_{i\overline{j}}.u^j.e^i=h_{ij}.u^{\overline{j}}.e^i=u(h_u)=0\text{ ,%
} \\ 
h_{ij}=g_{ij}-\frac 1e.u_i.u_j\text{ ,} \\ 
\\ 
h_{i\overline{j}}=f^k\text{ }_j.h_{ik}\text{ , }u^{\overline{j}}=f^j\text{ }%
_k.u^k\text{ , }u_i=g_{ik}.u^{\overline{k}}=g_{i\overline{k}}.u^k\text{ ,}
\\ 
f^i\text{ }_j=S(e^i,e_j)=S(e_j,e^i)=e^i(e_j)\in C^r(M)\text{ ,}
\end{array}
\label{14}
\end{equation}

\textit{Remark}. In $(\overline{L}_n,g)$-spaces (dim$\,M=n$) $S$ is the
contraction operator. In $(L_n,g)$-spaces [as special case of $(\overline{L}%
_n,g)$-spaces] $S=C:C(e^i,e_j)=g_j^i=e^i(e_j)$, $g_j^i=1$ for $i=j$ and $%
g_j^i=0$ for $i\neq j$.

The tensor $\sigma $ is the \textit{shear velocity tensor} (shear) which can
be written in the form 
\begin{equation}
\begin{array}{c}
\sigma =\frac 12\{h_u(\nabla _u\overline{g}-\pounds _u\overline{g})h_u-\frac
1{n-1}(h_u[\nabla _u\overline{g}-\pounds _u\overline{g}]).h_u\}=\sigma
_{ij}.e^i.e^j\text{ ,} \\ 
\\ 
\overline{g}=g^{ij}.e_i.e_j\text{ , }e_i.e_j=\frac 12(e_i\otimes
e_j+e_j\otimes e_i)\text{ ,}
\end{array}
\label{15}
\end{equation}
\begin{equation}
\sigma _{ij}=\frac 12\{h_{i\overline{k}}(g^{kl}\text{ }_{;m}u^m-\pounds
_ug^{kl})h_{\overline{l}j}-\frac 1{n-1}.h_{\overline{k}\overline{l}}(g^{kl}%
\text{ }_{;m}u^m-\pounds _ug^{kl}).h_{ij}\text{ ,}  \label{16}
\end{equation}

\noindent where 
\begin{equation*}
\begin{array}{c}
g(u)=g_{i\overline{k}}u^k.e^i=g_{ik}u^{\overline{k}}.e^i\text{ , }g_{i%
\overline{k}}=f^l\text{ }_k.g_{il}\text{ , }\overline{g}[g(u)]=u\text{ ,} \\ 
\\ 
g[\overline{g}(p)]=p\text{ , }p\in T^{*}(M)\text{ , }g^{ij}.g_{\overline{j}%
\overline{k}}=g_k^i\text{ , }g^{\overline{i}\overline{j}}.g_{jk}=g_k^i\text{
,} \\ 
\\ 
g_{\overline{j}\overline{k}}=f^l\text{ }_j.f^m\text{ }_k.g_{lm}\text{ , }g^{%
\overline{i}\overline{j}}=f^i\text{ }_k.f^j\text{ }_l.g^{kl}\text{ ,}
\end{array}
\end{equation*}

The tensor $\omega $ is the \textit{rotation velocity tensor} (rotation), 
\begin{equation}
\begin{array}{c}
\omega =h_u(k_a)h_u=h_u(s)h_u-h_u(q)h_u=S-Q=\omega _{ij}.e^i\wedge e^j\text{
,} \\ 
\\ 
s=\frac 12(u^k\text{ }_{;m}g^{ml}-u^l\text{ }_{;m}g^{mk})e_k\wedge e_l\text{
,} \\ 
\\ 
q=\frac 12(T_{mn}^kg^{ml}-T_{mn}^lg^{mk})u^n.e_k\wedge e_l\text{ ,} \\ 
\\ 
S=h_u(s)h_u=h_{i\overline{k}}.s^{kl}.h_{\overline{l}j}.e^i\wedge e^j\text{ , 
}Q=h_u(q)h_u\text{ ,} \\ 
\\ 
e^i\wedge e^j=\frac 12(e^i\otimes e^j-e^j\otimes e^i)\text{ , \thinspace
\thinspace \thinspace \thinspace \thinspace \thinspace \thinspace \thinspace
\thinspace \thinspace \thinspace \thinspace }e_k\wedge e_l=\frac
12(e_k\otimes e_l-e_l\otimes e_k)\text{ ,}
\end{array}
\label{17}
\end{equation}

The invariant $\theta $ is the \textit{expansion velocity} invariant
(expansion), 
\begin{equation}
\theta =\frac 12h_u[\nabla _u\overline{g}-\pounds _u\overline{g}]=\frac 12h_{%
\overline{i}\overline{j}}(g^{ij}\text{ }_{;k}u^k-\pounds _ug^{ij})\text{ .}
\label{18}
\end{equation}

In this way the notions of shear, rotation and expansion are generalized for 
$(\overline{L}_n,g)$-spaces. In analogous way (after some more complicated
computations) for the second covariant derivative $\nabla _u\nabla _u\xi $
notions such as shear acceleration, rotational acceleration and expansion
acceleration can be introduced in $\overline{V}_n$-spaces [(pseudo)
Riemannian spaces without torsion with contraction operator $S\neq C$], in $%
\overline{U}_n$-spaces [(pseudo) Riemannian spaces with torsion and with
contraction operator $S\neq C$] and in $(\overline{L}_n,g)$-spaces. These
notion can also be connected with the generalized deviation identity which
can be written in the form 
\begin{equation*}
\nabla _u\nabla _u\xi \equiv \frac 1e.g(u,\nabla _u\nabla _u\xi ).u+%
\overline{g}[h_u(\nabla _u\nabla _u\xi )]\text{ .} 
\end{equation*}

After straightforward calculations the orthogonal to $u$ part $\overline{g}[%
h_u(\nabla _u\nabla _u\xi )]$ of $\nabla _u\nabla _u\xi $ can be represented
in the form \cite{Manoff-05} 
\begin{equation}
\begin{array}{c}
\overline{g}[h_u(\nabla _u\nabla _u\xi )]=\overline{g}(h_u)[\frac le.\nabla
_ua-\nabla _{\pounds _\xi u}u-\nabla _u(\pounds _\xi u)+T(\pounds _\xi u,u)]+
\\ 
\\ 
+\overline{g}[_sD(\xi )+W(\xi )+\frac 1{n-1}.U.h_u(\xi )]\text{ ,}
\end{array}
\label{19}
\end{equation}

\noindent or in a given basis (in index form) 
\begin{equation}
\begin{array}{c}
g^{ij}.h_{\overline{j}\overline{k}}(\xi ^k\text{ }%
_{;l}u^l)_{;m}u^m=g^{ij}.h_{\overline{j}\overline{k}}[\frac le.a^k\text{ }%
_{;l}u^l-u^k\text{ }_{;l}\pounds _\xi u^l-(\pounds _\xi
u^k)_{;l}u^l+T_{mn}^k\pounds _\xi u^mu^n]+ \\ 
\\ 
+g^{i\overline{j}}(_sD_{jk}+W_{jk}+\frac 1{n-1}.U.h_{jk})\xi ^{\overline{k}}%
\text{ .}
\end{array}
\label{20}
\end{equation}

The tensor $_sD=_{sF}D_0-_{sT}D_0+_sM$ is the \textit{shear acceleration}
tensor (shear acceleration) constructed by three terms: the tensor $_{sF}D_0$
is the curvature- and torsion-free shear acceleration, the tensor $_{sT}D_0$
is the shear acceleration, induced by torsion, the tensor $_sM$ is the shear
acceleration, induced by curvature; the tensor $W=_FW_0-_TW_0+N$ is the 
\textit{rotation acceleration} tensor (rotation acceleration) which has also
three terms: the tensor $_FW_0$ is the curvature- and torsion-free rotation
acceleration, the tensor $_TW_0$ is the rotation acceleration, induced by
torsion, the tensor $N$ is the rotation acceleration, induced by curvature;
the invariant $U=_FU_0-_TU_0+I$ is the \textit{expansion acceleration}
invariant (expansion acceleration) with the three terms: the invariant $%
_FU_0 $ is the curvature- and torsion-free expansion acceleration, the
invariant $_TU_0$ is the expansion acceleration, induced by torsion, the
invariant $I$ is the expansion acceleration, induced by curvature [this term
appears as a generalization of the Raychaudhuri identity \cite{Kramer} for $(%
\overline{L}_n,g)$-spaces].

By means of different representations of the generalized deviation identity
possibilities can be considered for writing down theoretical schemes in
gravitational theories (and particular in the ETG) for construction of
gravitational wave detectors.

\section{Deviation equation of Synge and Schild}

The deviation equation of Synge and Schild in $(\overline{L}_{n},g)$-spaces
can be obtained from the generalized deviation identity by means of the
additional condition $\pounds _{\xi }u=0$ or $\pounds _{\xi }u^{i}=0$ in the
form 
\begin{equation}
\nabla _{u}\nabla _{u}\xi =[R(u,\xi )]u+\nabla _{\xi }a-\nabla _{u}[T(\xi
,u)]\text{ , \ \ \ \ \ \ \ \ \ }a=\nabla _{u}u\text{ ,}  \label{4.1}
\end{equation}

\noindent or in a arbitrary basis (or in index form) 
\begin{equation}
(\xi ^i\text{ }_{;j}u^j)_{;k}u^k=R_{klj}^iu^ku^l\xi ^j+a^i\text{ }_{;j}\xi
^j-(T_{kl}^i\xi ^ku^l)_{;j}u^j\text{ .}  \label{4.2}
\end{equation}

At the same time the conditions 
\begin{equation}
\begin{array}{c}
\nabla _{u}\xi =\nabla _{\xi }u-T(\xi ,u)\text{ \ \ \ or \ \ \ \ }\xi ^{i}%
\text{ }_{;j}u^{j}=u^{i}\text{ }_{;j}\xi ^{j}-T_{kl}^{i}\xi ^{k}u^{l}\text{ ,%
} \\ 
\\ 
\pounds _{\xi }a=[\pounds \Gamma (\xi ,u)]u\text{ \ \ \ or \ \ \ }\pounds
_{\xi }\Gamma _{jk}^{i}.u^{j}u^{k}=\pounds _{\xi }a^{i}\text{ ,}
\end{array}
\label{4.3}
\end{equation}

\noindent are fulfilled.

The way of getting the deviation equation of Synge and Schild gives the
possibility for proving the following proposition:

\textbf{Proposition 1.} Every vector field $\xi $, which satisfies the
equation $\pounds _{\xi }u=0$ ($\pounds _{\xi }u^{i}=0$) for an arbitrary
vector field $u$ is a solution of the deviation equation of Synge and Schild.

Proof: There are at least two ways for proving this proposition:

1. The proof follows immediately from the generalized identity and the
condition $\pounds _\xi u=0$.

2. From the condition $\pounds _\xi u=0$ and after covariant differentiation
along $u$ of the expression for $\nabla _u\xi $ (s. above) the deviation
equation follows.

\textbf{Corollary. }The condition $\pounds _\xi u=0$ is a ''first integral''
for the deviation equation of Synge and Schild (for arbitrary vector field $%
u $).

\textit{Remark}\textbf{. }Under ''first integral'' here one can define a
quantity whose covariant derivative along an arbitrary vector field $u$
leads to the deviation equation of a concrete type (here of Synge and
Schild).

\textbf{Proposition 2.} The necessary and sufficient condition for the
existence of the deviation equation of Synge and Schild is the condition $%
\pounds _\xi a=[\pounds \Gamma (\xi ,u)]u$ or $\pounds _\xi a^i=\pounds _\xi
\Gamma _{kl}^i.u^ku^l$.

Proof: (a) Necessity: From the generalized deviation identity and the
deviation equation of Synge and Schild the condition follows.

(b) Sufficiency: From the condition and the generalized deviation identity
the deviation equation of Synge and Schild follows.

\textit{Remark}\textbf{.} In finding out deviation equations different
authors used only sufficient (or ''first integrals'') conditions for these
equations (like those in proposition 2.). They don't take into account that
the obtained equations can fulfill also other sufficient conditions than the
considered one (s. for example \cite{Manoff-d}, \cite{Swaminarayan}).

In a $(\overline{L}_n,g)$-space, the second covariant derivative of a vector
field $\xi $ along a non-isotropic (non-null) vector field $u$ can be
written in two parts: the one is collinear to $u$, the other is orthogonal
to the vector field $u$. The second term can be interpreted as a relative
acceleration between two points, lying on a hyper-surface orthogonal to the
vector field $u$. Since the (infinitesimal) deviation vector has also to lie
on this hyper surface, then in this case $\xi $ has to obey the condition 
\begin{equation}
g(\xi ,u)=0\text{ ,}  \label{4.4}
\end{equation}

\noindent or $\xi $ has to be in the form 
\begin{equation}
\xi _{\perp }=\overline{g}[h_{u}(\xi )]=g^{ik}h_{\overline{k}\overline{l}%
}\xi ^{l}.e_{i}\text{ , \ \ \ \ \ \ \ \ \ \ \ }g(\xi _{\perp },u)=0\text{ .}
\label{4.5}
\end{equation}

\textbf{Definition 1.} The deviation equation which is obtained for 
\begin{equation*}
\overline{g}[h_{u}(\nabla _{u}\nabla _{u}\xi )]\text{ \ \ \ or for \ \ }%
h_{u}(\nabla _{u}\nabla _{u}\xi ) 
\end{equation*}
under the conditions 
\begin{equation}
\pounds _{\xi _{\perp }}u=0\text{ , \ \ \ \ \ \ \ \ \ \ }g(u,\xi _{\perp })=0%
\text{ , \ \ \ \ \ \ \ \ }\xi _{\perp }=\overline{g}[h_{u}(\xi )]\text{ ,}
\label{4.6}
\end{equation}

\noindent is called \textit{projective deviation equation of Synge and Schild%
}.

After some calculations, it follows from the form of $\overline{g}[%
h_u(\nabla _u\nabla _u\xi $ $)]$ that this equation can be written in the
form \cite{Manoff-05} 
\begin{equation}
\overline{g}[h_u(\nabla _u\nabla _u\xi _{\perp })]=\overline{g}[A(\xi
_{\perp })]=\overline{g}[_sD(\xi _{\perp })]+\overline{g}[W(\xi _{\perp })%
]+\frac 1{n-1}.U.\xi _{\perp }\text{ ,}  \label{4.7}
\end{equation}

\noindent or in index form 
\begin{equation}
\begin{array}{c}
g^{ij}h_{\overline{j}\overline{k}}(\xi _{\perp ;l}^ku^l)_{;m}u^m=g^{ij}A_{%
\overline{j}\overline{k}}\xi _{\perp }^k= \\ 
\\ 
=g^{ij}(_sD_{\overline{j}\overline{k}}+W_{\overline{j}\overline{k}})\xi
_{\perp }^k+\frac 1{n-1}.U.\xi _{\perp }^i\text{ ,}
\end{array}
\label{4.8}
\end{equation}

\noindent where 
\begin{equation*}
\begin{array}{c}
\xi _{\perp }^{k}=g^{kl}.h_{\overline{l}\overline{m}}.\xi ^{m}\text{ , \ \ \
\ \ \ \ \ }h_{u}(\xi _{\perp })=h_{u}(\overline{g})h_{u}(\xi )=h_{u}(\xi )%
\text{ ,} \\ 
\\ 
\overline{g}[h_{u}(\xi _{\perp })]=\overline{g}[h_{u}(\xi )]=\xi _{\perp }%
\text{ .}
\end{array}
\end{equation*}

The projective deviation equation can also be written in an equivalent form 
\begin{equation}  \label{4.9}
h_u(\nabla _u\nabla _u\xi _{\perp })=_sD(\xi _{\perp })+W(\xi _{\perp
})+\frac 1{n-1}.U.g(\xi _{\perp })
\end{equation}

Every vector field $\xi _{\perp }$ [for an arbitrary non-isotropic
(non-null) vector field $u$] which fulfills the conditions $\pounds _{\xi
_{\perp }}u=0$, $\xi _{\perp }=\overline{g}[h_u(\xi )]$, is a solution of
the projective deviation equation of Synge and Schild. Therefore, the
solution of equation $\pounds _{\xi _{\perp }}u=0$ (or $\pounds _u\xi
_{\perp }=0$) for a vector field $\xi _{\perp }(x^k)$ and a given vector
field $u(x^k)$ is also a solution of the projective deviation equation. It
follows in this case that, if the components of the vector field $\xi =\xi
^i.e_i=\xi ^k.\partial _k$ should be solutions of a homogeneous (or
non-homogeneous) oscillator equation, then an additional equation for the
vector field $u$ has to be proposed, which could lead to such properties of $%
\xi $.

A deviation equation under the same conditions $\pounds _{\xi _{\perp }}u=0$%
, $\xi _{\perp }=\overline{g}[h_u(\xi )]$, can also be written for the
square of $\xi _{\perp }$, i.e. for $g(\xi _{\perp ,}\xi _{\perp })=L^2\neq
0.$ If the vector field $u$ is considered as a time like vector field which
is orthogonal to $\xi _{\perp }$, then $\xi _{\perp }$ could be interpreted
as a space like vector field which length is considered as the length of a
material object or the length of the distance between two particles, lying
on an orthogonal to $u$ hypersurface.

By means of the relations 
\begin{equation}
\begin{array}{c}
\nabla _u\xi _{\perp }=_{rel}v+\overline{g}(\nabla _u\xi )(\xi )+(\nabla _u%
\overline{g})(h_u(\xi ))\text{ ,} \\ 
\\ 
_{rel}v=\overline{g}[h_u(\nabla _u\xi )]=g^{ij}h_{\overline{j}\overline{k}%
}\xi ^k\text{ }_{;l}u^l.e_i\text{ ,} \\ 
\\ 
\nabla _u\nabla _u\xi _{\perp }=_{rel}a+2\overline{g}(\nabla _uh_u)(\nabla
_u\xi )+\overline{g}(\nabla _u\nabla _uh_u)(\xi )+2(\nabla _u\overline{g}%
)g(_{rel}v)+ \\ 
\\ 
+2(\nabla _u\overline{g})(\nabla _uh_u)(\xi )+(\nabla _u\nabla _u\overline{g}%
)g(\xi _{\perp })\text{ ,} \\ 
\\ 
_{rel}a=\overline{g}[h_u(\nabla _u\nabla _u\xi )]=g^{ij}h_{\overline{j}%
\overline{k}}(\xi ^k\text{ }_{;l}u^l)_{;m}u^m.e_i\text{ ,} \\ 
\\ 
(\nabla _u\overline{g})g(_{rel}v)=(\nabla _u\overline{g})(g)(_{rel}v)=g^{ij}%
\text{ }_{;k}u^kg_{\overline{j}\overline{l}}._{rel}v^l.e_i\text{ ,}
\end{array}
\label{4.10}
\end{equation}

\noindent the deviation equation for $L^2$ can be obtained in the form 
\begin{equation}
\begin{array}{c}
u(uL^2)=2[g(\xi _{\perp },_{rel}a)+2g(\xi _{\perp },\overline{g}(\nabla
_uh_u)(\nabla _u\xi ))+g(\xi _{\perp },\overline{g}(\nabla _u\nabla
_uh_u)(\xi ))+ \\ 
\\ 
+2g(\xi _{\perp },(\nabla _u\overline{g})g(_{rel}v))+2g(\xi _{\perp
},(\nabla _u\overline{g})(\nabla _uh_u)(\xi ))+g(\xi _{\perp },(\nabla
_u\nabla _u\overline{g})g(\xi _{\perp }))]+ \\ 
\\ 
+2[g(_{rel}v,_{rel}v)+2g(_{rel}v,\overline{g}((\nabla _uh_u)(\xi ))+g(%
\overline{g}(\nabla _uh_u)(\xi ),\overline{g}((\nabla _uh_u)(\xi ))+ \\ 
\\ 
+2g(_{rel}v,(\nabla _u\overline{g})g(\xi _{\perp }))+2g(\overline{g}(\nabla
_uh_u)(\xi ),(\nabla _u\overline{g})g(\xi _{\perp }))+ \\ 
\\ 
+g((\nabla _u\overline{g})g(\xi _{\perp }),(\nabla _u\overline{g})g(\xi
_{\perp }))]+ \\ 
\\ 
+4(\nabla _ug)(\xi _{\perp },\nabla _u\xi _{\perp })+(\nabla _u\nabla
_ug)(\xi _{\perp },\xi _{\perp })\text{ .}
\end{array}
\label{4.11}
\end{equation}

For $\overline{U}_n$- and $\overline{V}_n$-spaces ($\nabla _ug=0$ for $%
\forall u\in T(M)$) this equation will have the form 
\begin{equation}
\begin{array}{c}
u(uL^2)=2[g(\xi _{\perp },_{rel}a)+2g(\xi _{\perp },\overline{g}(\nabla
_uh_u)(\nabla _u\xi ))+ \\ 
\\ 
+g(\xi _{\perp },\overline{g}(\nabla _u\nabla _uh_u)(\xi
))]+2[g(_{rel}v,_{rel}v)+ \\ 
\\ 
+2g(_{rel}v,\overline{g}(\nabla _uh_u)(\xi ))+g(\overline{g}(\nabla
_uh_u)(\xi ),\overline{g}(\nabla _uh_u)(\xi ))]\text{ .}
\end{array}
\label{4.12}
\end{equation}

If the additional condition (parallel transport of $h_u$ along $u$) 
\begin{equation}
\nabla _uh_u=0  \label{4.13}
\end{equation}

\noindent is required, then the equation for $L^2$ will have the form 
\begin{equation}
u(uL^2)=2[g(\xi _{\perp },_{rel}a)+g(_{rel}v,_{rel}v)]\text{ ,}  \label{4.14}
\end{equation}

\noindent or in index form 
\begin{equation}
((L^2)_{,i}u^i)_{,j}u^j=2(g_{\overline{k}\overline{l}}\xi _{\perp
}^k._{rel}a^l+g_{\overline{k}\overline{l}}._{rel}v^k._{rel}v^l)\text{ .}
\label{4.15}
\end{equation}

\textit{Remark}. If $u=d/ds$ is a tangent vector at a curve $x(s)$ then $%
u(uL^2)=d^2L^2/ds^2$.

The next task is to consider the deviation equation for $L^2$ for
auto-parallel ($\nabla _uu=a=0$), non-isotropic (non-null) $(g(u,u)=e\neq 0)$
and normalized $(e=const.\neq 0)$ vector field $u$.

\section{Projective deviation equation of Synge and Schild for $L^2$ in the
case of auto-parallel vector field $u$ in $\overline{U}_n$- and $\overline{V}%
_n$-spaces}

If the condition for auto parallelism is given for the vector field $u$,
i.e. if 
\begin{equation}  \label{4.16}
\nabla _uu=a=u^i\text{ }_{;j}u^j.e_i=a^i.e_i=0\text{ ,}
\end{equation}

\noindent then by means of the expression for $\nabla _uh_u$ in ($\overline{L%
}_n,g$)-spaces 
\begin{equation}
\begin{array}{c}
\nabla _uh_u=\nabla _ug+\frac 1e\{\frac 1e(ue).g(u)\otimes g(u)-[g(a)\otimes
g(u)+g(u)\otimes g(a)]- \\ 
\\ 
-[(\nabla _ug)(u)\otimes g(u)+g(u)\otimes (\nabla _ug)(u)]\}
\end{array}
\label{4.17}
\end{equation}

\noindent the following proposition can be proved for the case of $\overline{%
U}_n$-spaces [$\nabla _vg=0$ for $\forall v\in T(M)$]:

\textbf{Proposition 3.} For a non-isotropic, normalized and auto parallel
vector field $u$ in $\overline{U}_n$-space the condition for $L^2=g(\xi
_{\perp },\xi _{\perp })$%
\begin{equation}  \label{4.20}
u(uL^2)=2[g(\xi _{\perp },_{rel}a)+g(_{rel}v,_{rel}v)]
\end{equation}

\noindent is fulfilled.

Proof: The last condition follows immediately from the expression for $%
u(uL^2)$ and the condition $\nabla _uh_u=0$ (which is fulfilled in this
case).

Let us now use the representation for $\nabla _u\xi $ by means of the
kinematic characteristics $d,\sigma ,\omega ,\theta $ and for $\nabla
_u\nabla _u\xi _{\perp }$ by means of the kinematic characteristics $A,$ $%
_sD,$ $W,$ $U$ and their structure under the conditions of proposition 3.
Then, the last expression for $u(uL^2)$ can be written in the form 
\begin{equation}
u(uL^2)-\frac 2{n-1}.U.L^2=2[_sD(\xi _{\perp },\xi _{\perp })+\overline{g}%
(d(\xi _{\perp }),d(\xi _{\perp }))]\text{ ,}  \label{4.21}
\end{equation}

\noindent where 
\begin{equation*}
\begin{array}{c}
_sD(\xi _{\perp },\xi _{\perp })=(\xi _{\perp })(_sD(\xi _{\perp }))=_sD_{%
\overline{k}\overline{l}}\xi _{\perp }^k.\xi _{\perp }^l\text{ ,} \\ 
\\ 
d(\xi _{\perp })=d_{k\overline{l}}\xi _{\perp }^l.e^k\text{ ,}
\end{array}
\end{equation*}

\noindent and the following relations are fulfilled 
\begin{equation}
\begin{array}{c}
g(_{rel}v,_{rel}v)=(_{rel}v)^2=\overline{g}(d(\xi _{\perp }),d(\xi _{\perp
}))\text{ ,} \\ 
\\ 
g(\xi _{\perp },_{rel}a)=g(\xi _{\perp },\overline{g}[_sD(\xi _{\perp })%
])+\frac 1{n-1}.U.L^2\text{ ,} \\ 
\\ 
g(\xi _{\perp },\overline{g}[W(\xi _{\perp })])=0\text{ .}
\end{array}
\label{4.22}
\end{equation}

If we use the explicit form of $d=\sigma +\omega +\frac 1{n-1}.\theta .h_u$
and introduce the following abbreviations 
\begin{equation}
\begin{array}{c}
\lambda =-\frac 2{n-1}\{\overline{g}[\sigma (\overline{g})\sigma ]+\overline{%
g}[\omega (\overline{g})\omega ]+\theta ^{.}+\frac 2{n-1}.\theta ^2\}\text{ ,%
} \\ 
\\ 
_sD(\xi _{\perp },\xi _{\perp })=D^2\text{ , }\theta ^{.}=u\theta
=u^i\partial _i\theta =u^j.e_j\theta \text{ ,} \\ 
\\ 
\sigma (\xi _{\perp })=\delta \text{ , }\omega (\xi _{\perp })=\eta \text{ ,}
\\ 
\\ 
\frac{2\theta }{n-1}.\sigma (\xi _{\perp },\xi _{\perp })=\frac{2\theta }{n-1%
}.(\xi _{\perp })(\sigma (\xi _{\perp }))=\sigma ^2\text{ ,} \\ 
\\ 
(\delta +\eta )^2=\delta ^2+2\delta \eta +\eta ^2=\overline{g}(\delta
,\delta )+2\overline{g}(\delta ,\eta )+\overline{g}(\eta ,\eta )\text{ ,} \\ 
\\ 
L^2=y\text{ , }u=\frac d{ds}=(dx^i/ds).\partial _i=u^i.\partial _i\text{ ,}
\end{array}
\label{4.23}
\end{equation}

\noindent then after some computations we can obtain the equation for $%
u(uL^2)=d^2L^2/ds^2$ (for $u=d/ds$) in the form 
\begin{equation}
\frac{d^2y}{ds^2}+\lambda (s).y=f(s)\text{ ,}  \label{4.24}
\end{equation}

\noindent where $y=y(x^k(s))=y(s)$ , $x^k=x^k(s)$ , and 
\begin{equation*}
\begin{array}{c}
f(s)=2[D^2+(\delta +\eta )^2+\sigma ^2]\text{ ,} \\ 
\\ 
D^2=D^2(x^k(s))=D^2(s)\text{ , }\delta =\delta (x^k(s))\text{ , }\eta =\eta
(x^k(s))\text{ , }\sigma =\sigma (x^k(s))\text{ .}
\end{array}
\end{equation*}

The explicit forms of $\lambda (s)$ and $f(s)$ determine the explicit form
of the equation for $y$ and therefore its solutions as well.

It is worth to mention that the explicit form of $\lambda $ and $f$ can be
found after solving the equations for the vector fields $u$ and $\xi $: $%
\nabla _uu=0$ , $\pounds _\xi u=0$ under the additional conditions $%
g(u,u)=e=const.\neq 0$, $g(u,\xi )=0$.

From the form of the equation for $y$ one can draw a conclusion that the
equation for $y$ could have a form of oscillator equation (homogeneous or
non-homogeneous) under the condition 
\begin{equation*}
\lambda (s)=\lambda _0=const.\neq 0\text{ ,} 
\end{equation*}

\noindent which is a very special case, requiring additional discussions.

In the case of $\overline{U}_{n}$-space admitting non-isotropic (non-null) ,
auto-parallel and normalized vector field $u$ with shear $\sigma =0$ and
rotation $\omega =0$%
\begin{equation}
\lambda =-\frac{2}{n-1}(\theta ^{.}+\frac{2}{n-1}.\theta ^{2})\text{ , \ }%
D^{2}=0\text{ , \ \ }\delta =0\text{ , \ }\eta =0\text{ , \ }\sigma ^{2}=0%
\text{ ,}  \label{4.25}
\end{equation}

\noindent the equation for $y$ will have the form 
\begin{equation}
\begin{array}{c}
y^{..}=\{[g(s)]^2+g^{.}(s)\}.y\text{ ,\thinspace \thinspace \thinspace
\thinspace \thinspace \thinspace \thinspace \thinspace \thinspace \thinspace
\thinspace \thinspace \thinspace }y^{.}=\frac{dy}{ds}\text{ ,\thinspace
\thinspace \thinspace \thinspace \thinspace \thinspace \thinspace \thinspace
\thinspace \thinspace \thinspace \thinspace \thinspace \thinspace \thinspace
\thinspace \thinspace \thinspace \thinspace \thinspace \thinspace \thinspace
\thinspace \thinspace }y^{..}=\frac{d^2y}{ds^2}\text{ ,} \\ 
\\ 
g(s)=\frac 2{n-1}.\theta \text{ , \thinspace \thinspace \thinspace
\thinspace \thinspace \thinspace \thinspace \thinspace \thinspace \thinspace
\thinspace \thinspace \thinspace \thinspace \thinspace \thinspace \thinspace
\thinspace \thinspace }g^{.}(s)=\frac 2{n-1}.\theta ^{.}\text{ .}
\end{array}
\label{4.26}
\end{equation}

One solution of the last equation has been found by Ielchin \cite{Kamke} in
the form 
\begin{equation}  \label{4.27}
y=\exp \int g(s)ds\text{ .}
\end{equation}

In the case of $\overline{V}_n$-space ($n=4$) under the conditions $\pounds
_{\xi _{\perp }}u=0$, $g(\xi _{\perp },u)=0$ , and the conditions for $u$ to
be non-isotropic, normalized and auto parallel vector field for $L^2=g(\xi
_{\perp },\xi _{\perp })$ the following deviation equation can be obtained 
\begin{equation}  \label{4.28}
u(uL^2)-\frac 2{n-1}.I.L^2=2[_sM(\xi _{\perp },\xi _{\perp })+ \overline{g}%
(d(\xi _{\perp }),d(\xi _{\perp }))]\text{ ,}
\end{equation}

\noindent where $U=I$ ($_FU=0$) , $_sD=\,_sM$.

The last equation for $L^2$ (if $u=\frac d{ds}$ , $\nabla _u=\frac D{ds}$)
can be written therefore in the form 
\begin{equation*}
\frac{d^2y}{ds^2}+\lambda (s).y=f(s)\text{ ,} 
\end{equation*}

\noindent where 
\begin{equation}
\begin{array}{c}
\lambda (s)=-\frac 2{n-1}(I+\frac 1{n-1}.\theta ^2)\text{ ,} \\ 
\\ 
I=\overline{g}[\sigma (\overline{g})\sigma ]+\overline{g}[\omega (\overline{g%
})\omega ]+\theta ^{.}+\frac 1{n-1}.\theta ^2\text{ ,} \\ 
\text{(Raychaudhuri identity),} \\ 
\\ 
f(s)=2[M^2+(\delta +\eta )^2+\sigma ^2]\text{ ,} \\ 
\\ 
M^2=_sM(\xi _{\perp },\xi _{\perp })=_sM_{\overline{k}\overline{l}}.\xi
_{\perp }^k.\xi _{\perp }^l\text{ .}
\end{array}
\label{4.29}
\end{equation}

For $\overline{V}_n$-spaces with $Ricci=0$ ($R_{ij}=0$) the equation for $y$
takes the form 
\begin{equation}  \label{4.30}
y^{..}-\frac 2{(n-1)^2}.\theta ^2.y=f(s)\text{ , }y^{.}=\frac{dy }{ds}\text{
.}
\end{equation}

If for such type of spaces the conditions $\sigma =0$, $\omega =0$ are
fulfilled, then 
\begin{equation}
_{s}M=0\text{, \ \ \ \ }\delta =0\text{, \ \ \ \ \ }\eta =0\text{, \ \ \ \ \ 
}\sigma ^{2}=0\text{, \ \ \ \ \ \ \ }f(s)=0\text{ .}  \label{4.31}
\end{equation}

The equation for $y$ will have the form 
\begin{equation}  \label{4.32}
y^{..}=\frac 2{(n-1)^2}.\theta ^2.y\text{ ,}
\end{equation}

\noindent which by means of the substitutions \cite{Kamke} $y^{.}=y.v(s)$
can be transformed in a Riccati equation 
\begin{equation}
v^{.}+v^2=\frac 2{(n-1)^2}.\theta ^2\text{ .}  \label{4.33}
\end{equation}

If $v(s)$ is one solution of this equation, then the solutions for $y$ are
solutions of a 1st order linear differential equation 
\begin{equation}
y^{.}-v(s).y=C.\exp (-\int v.ds)\text{ , \ \ \ \ \ \ \ \ \ }C=const.\text{ .}
\label{4.34}
\end{equation}

For the special case, when $\theta =\theta _{0}=const.\neq 0$, the equation $%
y^{..}=\frac{2}{(n-1)^{2}}.\theta ^{2}.y$ has the form 
\begin{equation}
y^{..}+\lambda _{0}.y=0\text{ , \ \ \ \ \ \ \ \ \ }\lambda _{0}=-\frac{2}{%
(n-1)^{2}}.\theta _{0}^{2}<0\text{ ,}  \label{4.35}
\end{equation}

and solutions of a type 
\begin{equation}
y=L^{2}=a.\cosh (\frac{\sqrt{2}}{n-1}.\theta _{0}.s)+b.\sinh (\frac{\sqrt{2}%
}{n-1}.\theta _{0}.s)\text{ , \ \ }a,b=\text{const .}  \label{4.36}
\end{equation}

Therefore, a deviation equation with non-isotropic, normalized and
auto-parallel (time like) vector field $u$ for the square of a non-isotropic
orthogonal (space like) to $u$ vector field $\xi $ can be considered as an
eventual candidate for the theoretical scheme of gravitational waves
detectors because such equations of this type could have, under certain
conditions in $\overline{U}_n$- and $\overline{V}_n$-spaces, the form of an
oscillator equation.

\textit{Remark}\textbf{.} The deviation equations of Synge and Schild have
equivalent forms in $(L_n,g)$- and in $(\overline{L}_n,g)$-spaces. These
equations are different in their forms if written in a given basis \cite
{Manoff-4}. This allows a general consideration for the both types of spaces.

\section{Conclusions}

The deviation equation of Synge and Schild and its corresponding projective
deviation equation in $(\overline{L}_n,g)$-spaces can be considered as a
corollary of the equation $\pounds _\xi u=0$ ($\pounds _u\xi =0$) for a
vector field $\xi $ and an arbitrary vector field $u$. The last equation
appears only as a sufficient, but not necessary condition for the existence
of the deviation equation of Synge and Schild, which, therefore, allows
other ''first integrals'' as well.

A deviation equation can also be considered for the square $L^2$ of a
non-isotropic (space like) vector field $\xi $, which equation appears in
fact as equation for an invariant, carrying information about the length of
this vector field. In the case of non-isotropic, normalized and
auto-parallel vector fields $u$ in $\overline{U}_n$-spaces and $\overline{V}%
_n$-spaces this equation could have the form of an oscillator equation under
certain conditions. This fact could be explored when theoretical schemes for
gravitational waves detectors are considered in a fixed gravitational theory
in $(\overline{L}_n,g)$-spaces, $\overline{U}_n$-spaces and $\overline{V}_n$%
-spaces.

\begin{center}
\textbf{Acknowledgments}
\end{center}

This work is supported in part by the National Science Foundation in
Bulgaria.

\end{document}